\documentclass[prl,aps,10pt,longbibliography,superscriptaddress,twocolumn]{revtex4-2}
\usepackage{amsmath}
\usepackage{amssymb}
\usepackage{amsthm}
\usepackage{amsfonts}
\usepackage{enumerate}
\usepackage{latexsym}
\usepackage{bm}
\usepackage{graphicx}
\usepackage{color}
\usepackage[dvipsnames]{xcolor}
\usepackage[normalem]{ulem}
\usepackage[colorlinks]{hyperref}
\hypersetup{
  colorlinks,
  citecolor=Violet,
  linkcolor=Red,
  urlcolor=Blue}
\usepackage{cancel}
\usepackage{mathrsfs}
\usepackage{braket}
\usepackage{float}

\begin{document}

\title{
Quantum advantage by designing beneficial dissipation from lossless dynamics
}

\author{Zeng-Zhao Li}
\email{lizengzhao@iqasz.cn}
\affiliation{International Quantum Academy, Shenzhen, 518048, China}

\author{Xiao Xue}
\email{xiao.xue@ustc.edu.cn}
\affiliation{International Quantum Academy, Shenzhen, 518048, China}
\affiliation{Hefei National Laboratory, University of Science and Technology of China, Hefei 230088, China}

\begin{abstract}
Passive loss removes probability from a system and therefore seems unable to improve an unconditional target. This intuition is exact for classical population dynamics under sink-only removal: passive killing obeys a finite-strength monotonicity principle even after optimization over the remaining controls. Here we show that coherent quantum dynamics can violate this classical monotonicity because coherent contributions to the final target can interfere destructively. We derive a trajectory-level criterion that determines when such a detrimental contribution is physically accessible to passive loss and accumulates into a net unconditional benefit. The criterion is encoded in a response operator constructed entirely from the lossless dynamics: its spectrum determines whether weak passive loss can help, while its principal eigenvector identifies the optimal environmental mode. Applying this criterion to vibronically assisted transfer, we identify a beneficial mode formed by coherently combining two individually detrimental local loss directions. The predicted mode remains near-optimal at finite loss, producing an \(8.73\%\) enhancement over the independently optimized lossless benchmark, compared with \(9.52\%\) under full finite-loss optimization. Our results establish how coherent quantum dynamics can evade classical passive-loss monotonicity and provide a general task-dependent principle for designing beneficial passive environments from lossless dynamics.
\end{abstract}

\date{\today}
\maketitle


Dissipation is usually associated with the loss of useful quantum resources, yet appropriately structured environments can also assist quantum dynamics. Noise and dephasing can enhance transport by suppressing destructive interference or releasing population from coherent trapping~\cite{Plenio2008,Mohseni2008,Rebentrost2009,Caruso2009,ZerahHarush2018,Maier2019}, while engineered reservoirs can prepare and stabilize selected quantum states, generate entanglement, steer quantum dynamics and implement autonomous quantum control~\cite{Poyatos1996,Diehl2008,Kraus2008,Verstraete2009,Barreiro2011,Lin2013,Shankar2013,Leghtas2013,Bermudez2013,Vuglar2018,Harrington2022}. 
In many such settings, however, the environment does more than simply remove probability: it redistributes population, modifies coherent pathways, or recycles excitations. 
Passive loss and non-Hermitian dynamics can also produce counterintuitive effects, including loss-induced transmission and revival, exceptional-point dynamics and loss-controlled quantum interference~\cite{Guo2009,Peng2014,Cerjan2016,Beder2024,Li2023,Zheng2025,Yu2026,LiXue2026}. 
These observations sharpen a more fundamental question: under what conditions can a purely passive sink improve an unconditional quantum target, and what distinguishes such an effect from its classical counterpart?

Recent trapped-ion experiments, for example, have demonstrated accelerated entanglement generation under non-Hermitian dynamics while revealing an accompanying reduction in the success probability~\cite{Yuan2026}. Such conditional acceleration need not increase the absolute target yield: the dynamical benefit must outweigh the survival cost.  
We therefore consider genuinely passive removal from the system of interest, with no population returned to the retained system and no conditioning on survival, and ask whether it can increase the unconditional target yield per unit time relative to the best lossless protocol within the same coherent control manifold. 
Classically, passive killing only attenuates non-negative trajectory contributions, precluding advantage even after reoptimization. Quantum advantage therefore requires a departure from this positivity. 

A natural candidate is quantum interference, but interference alone is not sufficient. Coherent dynamics can contain noncommuting forward- and backward-evolved operators, or locally signed responses, without providing a physically accessible loss channel that improves the task. 
What is required is an operational criterion that determines whether a beneficial passive environment exists, which accessible environmental mode realizes it, and whether its signed response accumulates favourably over the controlled trajectory. Such a criterion would turn qualitative interference arguments into a predictive principle for reservoir design.

\begin{figure*}[htbp]
\centering
 \includegraphics[width=1.8\columnwidth]{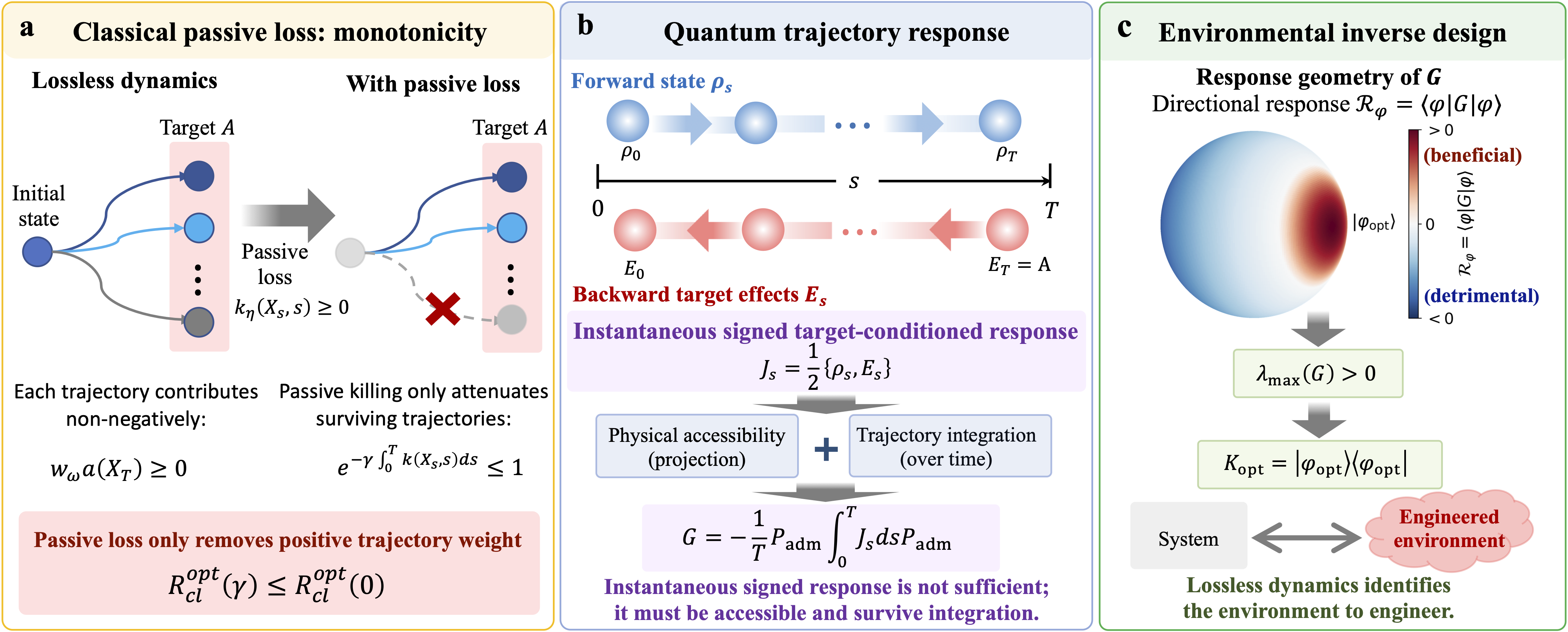} 
\caption{
{\bf Classical monotonicity, quantum trajectory response and environmental inverse design.}
{\bf a,} In a classical population process with sink-only removal, passive killing only attenuates non-negative trajectory contributions to the final target. Consequently, the optimized unconditional time-normalized target yield obeys $R_{\rm cl}^{\rm opt}(\gamma)\le R_{\rm cl}^{\rm opt}(0)$ for any passive-loss strength. 
{\bf b,} Quantum dynamics is governed instead by an instantaneous signed target-weighted response formed from the forward state \(\rho_s\) and the backward target effect \(E_s\), \(J_s=\{\rho_s,E_s\}/2\). An instantaneous signed response is not sufficient for operational advantage: it must survive physical accessibility (projection onto the admissible loss subspace) and trajectory integration over the controlled evolution. These operations define \(G=-(1/T)P_{\rm adm}\int_0^T J_s\,ds\,P_{\rm adm}\). 
{\bf c,} The spectrum of \(G\) provides an environmental design rule directly from the lossless trajectory. When any positive-semidefinite, trace-normalized loss operator supported on the admissible subspace can be implemented, beneficial weak passive loss exists iff \(\lambda_{\max}(G)>0\), and a principal eigenvector \(|\varphi_{\rm opt}\rangle\) specifies an optimal rank-one loss mode \(K_{\rm opt}=|\varphi_{\rm opt}\rangle\langle\varphi_{\rm opt}|\). Thus the lossless dynamics determines both whether weak passive dissipation can help and which environmental mode maximizes its weak-loss benefit.
}
\label{fig:Fig1}
\end{figure*}

Here we establish that beneficial passive dissipation is a task-dependent property of the lossless quantum trajectory. Projecting and integrating the target-weighted forward--backward response over the physically admissible loss space defines an operator \(G\) whose largest eigenvalue determines whether weak passive loss can help and whose principal eigenvector identifies the optimal environmental mode (Fig.~\ref{fig:Fig1}). 
We illustrate this general criterion first with a minimal Bell-state protocol that isolates physical accessibility and trajectory accumulation. We then test the resulting design principle in vibronically assisted transfer, where two local loss directions that are individually detrimental, and remain so under any incoherent mixing, form a beneficial environmental mode when combined coherently. Remarkably, the mode identified entirely from the lossless trajectory remains near-optimal at finite dissipation, producing an \(8.73\%\) enhancement compared with \(9.52\%\) under full optimization. The advantage persists over finite parameter regions and admits a microscopic auxiliary-reservoir realization.

\vspace{0.1cm}\noindent
{\bf Classical monotonicity and quantum possibility}\\
Consider a controlled continuous-time Markov process \(X_t\), with underlying control protocol \(\theta\), and let \(a(X_T)\geq0\) denote the terminal target contribution; for population transfer, it may be the indicator of target occupation at time \(T\).  
Passive loss is represented by a non-negative killing rate \(k_\eta(X_s,s)\), where \(\eta\) specifies its admissible spatial or temporal profile.  For fixed \((\theta,\eta)\), the unconditional target yield is \(Y_A^{\rm cl}(\gamma;\theta,\eta) = \mathbb E_{\theta}[a(X_T) e^{-\gamma\int_0^T k_\eta(X_s,s)\,ds}]\).
Here \(\mathbb E_{\theta}\) denotes the expectation over the lossless path measure.
Because both the terminal contribution and the survival factor are non-negative, passive killing can only reduce the contribution of every individual trajectory. Consequently \(Y_A^{\rm cl}(\gamma;\theta,\eta)\leq Y_A^{\rm cl}(0;\theta)\) for every admissible protocol and loss profile.
Defining the time-normalized target yield as \(R_{\rm cl}=Y_A^{\rm cl}/T\) and optimizing over the same underlying control family therefore gives
\begin{equation}
R_{\rm cl}^{\rm opt}(\gamma)
\leq
R_{\rm cl}^{\rm opt}(0),
\qquad \gamma\geq0 .
\label{eq:classical_nogo}
\end{equation}
This is a finite-strength result: reoptimizing either the control protocol or the passive-loss profile cannot reverse the pathwise ordering (Fig.~\ref{fig:Fig1}a; Supplementary Note~II). This sink-only setting is the classical counterpart of the irreversible, non-recycling passive loss considered below.

The quantum case differs fundamentally because coherent contributions to the final target need not be non-negative: an intermediate component can contribute constructively or destructively through quantum interference.
For a positive target operator \(A\), we define the unconditional target yield and its time-normalized counterpart as \(Y_A(T)=\operatorname{Tr}[A\rho(T)]\) and \(R(T) = Y_A(T) /T\), respectively. 
Here \(\rho(T)\) is subnormalized under passive loss; no renormalization by its trace is performed.
To resolve how intermediate components of the trajectory contribute to the final target, let \(\mathcal U_0(t,s)[X]\) denote the lossless dynamical map. We propagate the initial state forward and the target operator backward~\cite{Gammelmark2013}, i.e., \(\rho_s=\mathcal U_0(s,0)[\rho_0]\) and \(E_s=\mathcal U_0^\dagger(T,s)[A]\), respectively. 
Their Hermitian forward--backward response is the Jordan product
\begin{equation}
J_s=\frac12\{\rho_s,E_s\}.
\end{equation}
Although both \(\rho_s\) and \(E_s\) are positive, \(J_s\) need not be positive when they do not commute~\cite{GuoLuo2024}. For a normalized state \(|v\rangle\), the rank-one operator \(K_v=|v\rangle\langle v|\) specifies a possible passive-loss mode. A negative expectation value \(\langle v|J_s|v\rangle<0\) does not represent a negative probability; rather, it indicates a positive instantaneous first-order response of the final target yield to loss acting along this mode at time \(s\). Physically, removing amplitude from a coherent pathway can also modify its interference with the remaining pathways, potentially producing a favourable local response. Whether such a mode is physically accessible and yields a net benefit over the entire trajectory is determined below.

For a passive-loss channel with overall strength $\gamma\ge0$ and coupling operator $K\ge0$, where loss irreversibly removes population from the system of interest without recycling it, linearization about the lossless trajectory gives \(\frac{\partial Y_A}{\partial\gamma}|_{\gamma=0} = -\int_0^T ds\,\operatorname{Tr}(KJ_s)\). 
The passive-loss dynamics and the derivation of this response relation are given in Methods and Supplementary Note~I, respectively.
The environment is further restricted to a physically admissible loss subspace, with projector $P_{\rm adm}$. Defining
\begin{equation}
G=-\frac1T P_{\rm adm} \int_0^T ds\,J_s P_{\rm adm},
\label{eq:G}
\end{equation}
gives
\begin{equation}
\left.\frac{\partial R}{\partial\gamma}\right|_{\gamma=0} = \operatorname{Tr}(KG).
\end{equation}
Throughout, \(G\) acts on the physically admissible loss subspace \(\mathcal H_{\rm adm}\), and its eigenvalues are evaluated within this subspace (Fig.~\ref{fig:Fig1}b).  
If any positive-semidefinite, trace-normalized loss operator \(K\) supported on \(\mathcal H_{\rm adm}\) can be implemented, the maximum weak-loss response is \(\lambda_{\max}(G)\). Beneficial weak passive loss then exists if and only if \(\lambda_{\max}(G)>0\), with a principal eigenvector of \(G\) identifying an optimal rank-one environmental mode (Fig.~\ref{fig:Fig1}c). A positive eigenvalue represents a net unconditional gain, with the survival cost already included.

\begin{figure}[htbp]
\centering
 \includegraphics[width=0.99\columnwidth]{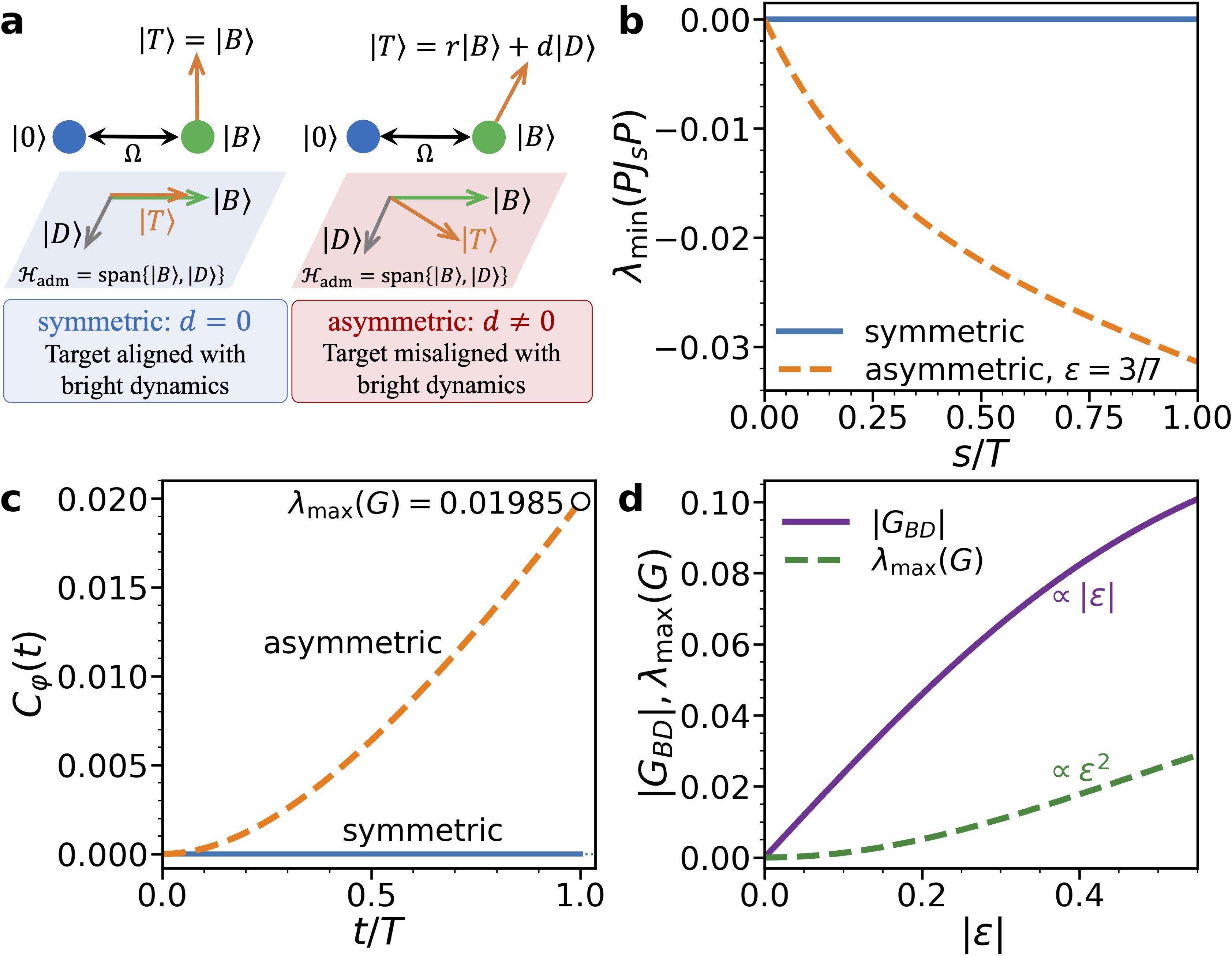} 
\caption{
{\bf Physical accessibility converts signed quantum response into a beneficial passive-loss direction.}
{\bf a,} In the bright--dark basis, coherent dynamics couples \(|0\rangle\) only to the bright state \(|B\rangle\), whereas the target is \(|T\rangle = r|B\rangle + d|D\rangle\). Symmetric driving gives \(d=0\), while drive asymmetry produces a target--dynamics misalignment \(d\neq0\). In both cases, the admissible loss space is \(\mathcal H_{\rm adm}=\mathrm{span}\{|B\rangle,|D\rangle\}\); asymmetry changes the target geometry rather than the admissible loss
subspace.
{\bf b,} Minimum eigenvalue \(\lambda_{\min}(PJ_sP)\) of the instantaneous response projected onto the admissible loss subspace, where \(P=P_{\rm adm}\). At symmetry, this eigenvalue remains zero throughout the lossless optimal trajectory, even though the full-space response \(J_s\) can be indefinite. For \(d\neq0\), the determinant of the projected response becomes strictly negative at intermediate times, implying \(\lambda_{\min}(PJ_sP)<0\) and making a negative instantaneous response direction physically accessible. The asymmetric curve uses \(\epsilon=(\Omega_1-\Omega_2)/(\Omega_1+\Omega_2)=3/7\). 
{\bf c,} Cumulative response \(C_\varphi(t)=-(1/T)\int_0^t ds\, \langle\varphi|J_s|\varphi\rangle\) along the optimal weak-loss mode \(|\varphi_{\rm opt}\rangle\). The asymmetric trajectory accumulates a positive response reaching \(C_{\varphi_{\rm opt}}(T) = \lambda_{\max}(G)=0.01985\), whereas the optimal symmetric admissible direction remains neutral.
{\bf d,} Perturbative opening of the beneficial channel from the symmetric limit. Drive asymmetry generates a bright--dark cross response linear in \(|\epsilon|\), \(|G_{BD} |\simeq0.23881| \epsilon|\), which mixes the neutral dark direction with the bright sector. The resulting beneficial
eigenvalue opens quadratically, \(\lambda_{\max}(G)\simeq 0.13505 \epsilon^2\).
}
\label{fig:Fig2}
\end{figure}

\vspace{0.1cm}\noindent
{\bf Physical accessibility of signed trajectory response}\\
A minimal Bell-state protocol illustrates two distinct requirements for beneficial passive loss: the instantaneous response \(J_s\) must have a negative direction within the physically admissible loss subspace, and the trajectory-integrated response \(G\) must have a positive eigenvalue corresponding to a fixed admissible environmental mode. 

We consider the three-state manifold $\{|0\rangle,|10\rangle,|01\rangle\}$, with $|0\rangle\equiv|00\rangle$, governed by \( H_0=\Omega_1|10\rangle\langle0| +\Omega_2|01\rangle\langle0| +\mathrm{H.c.}\)
Introducing the orthogonal bright and dark states
\( |B\rangle = (\Omega_1|10\rangle+\Omega_2|01\rangle)/\Omega\) and 
\( |D\rangle= (\Omega_2|10\rangle-\Omega_1|01\rangle)/\Omega\), with $\Omega=\sqrt{\Omega_1^2+\Omega_2^2}$, brings the Hamiltonian to \(H_0 =\Omega\bigl( |B\rangle\langle0| + |0\rangle\langle B|\bigr)\).
The coherent evolution therefore couples $|0\rangle$ exclusively to the bright state $|B\rangle$, while $|D\rangle$ is dynamically dark.

The target Bell state, $|T\rangle=(|10\rangle+|01\rangle)/\sqrt2$, has the decomposition \(|T\rangle=r|B\rangle+d|D\rangle\) with \(r = (\Omega_1+\Omega_2)/(\sqrt2\Omega)\) and \(d=(\Omega_2-\Omega_1)/(\sqrt2\,\Omega)\).
For symmetric driving, $\Omega_1=\Omega_2$ and hence $d=0$, so the target lies entirely along the dynamically bright direction. Drive asymmetry produces $d\neq0$, so the target acquires a dark-state component, creating a controlled target--dynamics misalignment (Fig.~\ref{fig:Fig2}a). 
We parameterize the drive asymmetry by \(\epsilon=(\Omega_1-\Omega_2)/(\Omega_1+\Omega_2)\), for which \(r=(1+\epsilon^2)^{-1/2}\) and \(d=-\epsilon(1+\epsilon^2)^{-1/2}=-\epsilon+O(\epsilon^3)\). 

Even at symmetry, the unprojected instantaneous response \(J_s\) can be indefinite, whereas physical loss acts only within \(\mathcal H_{\rm adm}=\operatorname{span}\{|B\rangle,|D\rangle\}\), with \(P\) denoting the projector onto this subspace. For \(d=0\), the minimum eigenvalue of the projected response remains zero throughout the lossless optimal trajectory, \(\lambda_{\min}(PJ_sP)=0\): the bright direction is detrimental and the dark direction neutral. Consequently, \(\lambda_{\max}(G)=0\) (Fig.~\ref{fig:Fig2}b).

Breaking the drive symmetry does not enlarge the admissible loss space, but changes the target geometry relative to the bright--dark basis. The determinant of the projected instantaneous response is \(\det(PJ_sP)=-\frac14 r^2d^2\sin^2(\Omega T)\sin^2(\Omega s)\leq0\). For the lossless-optimal trajectory used in Fig.~\ref{fig:Fig2}, this determinant is strictly negative when \(d\neq0\) and \(0<s<T\), implying \(\lambda_{\min}(PJ_sP)<0\) and making a negative instantaneous response direction physically accessible (Fig.~\ref{fig:Fig2}b). The analytical projected response and its eigenstructure are given in Supplementary Note~III.

Physical accessibility is still not sufficient. For a fixed environmental mode, the signed response must accumulate favourably along the same admissible direction over the trajectory. Here the loss mode is a normalized superposition \(|\varphi\rangle= c_B|B\rangle +c_D|D\rangle\), with \(|c_B|^2 +|c_D|^2=1\), specifying the rank-one coupling \(K_\varphi= |\varphi\rangle \langle\varphi|\).
Its cumulative response is \(C_\varphi(t)=-\frac1T \int_0^t ds\, \langle\varphi|J_s|\varphi\rangle\), whose terminal value satisfies \(C_\varphi(T)= \langle\varphi| G|\varphi\rangle\). The optimal mode \(|\varphi_{\rm opt}\rangle\) is a normalized eigenvector of \(G\) associated with its largest eigenvalue, so \(C_{\varphi_{\rm opt}}(T)=\lambda_{\max}(G)\). 
For \((\Omega_1,\Omega_2)=(1,0.4)\), corresponding to \(\epsilon=3/7\), this response reaches \(0.01985\), whereas the optimal admissible mode under symmetric driving remains neutral (Fig.~\ref{fig:Fig2}c).

Near symmetry, a small drive asymmetry generates a bright--dark integrated cross response \(|G_{BD}|\simeq 0.23881|\epsilon|\), while the dark diagonal response remains zero. This off-diagonal coupling to the detrimental bright direction therefore shifts the eigenvalue associated with the neutral dark direction only at second order, \(\lambda_{\max}(G)= 0.13505\,\epsilon^2 + O(\epsilon^4)\) (Fig.~\ref{fig:Fig2}d). The beneficial weak-loss response thus opens quadratically with asymmetry through coherent bright--dark mixing.

\begin{figure*}[t]
\centering
 \includegraphics[width=1.8
 \columnwidth]{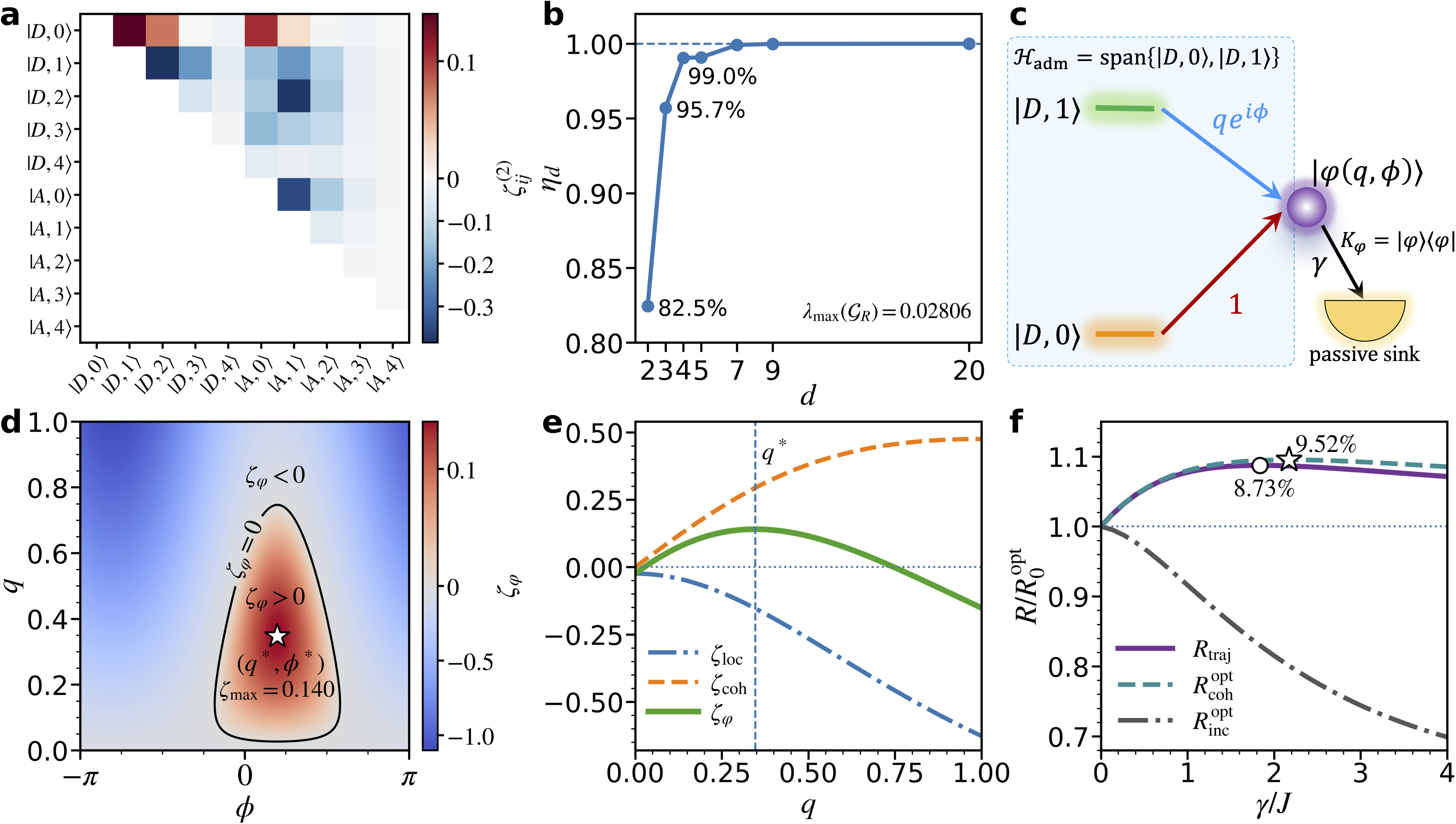} 
\caption{
{\bf Environmental coherence unlocks passive-dissipation advantage in vibronic quantum dynamics.}
{\bf a,} Pairwise weak-loss response $\zeta_{ij}^{(2)}$ obtained from the unrestricted lossless trajectory. Positive response is concentrated in a small set of low-lying vibronic sectors, with the donor pair $\{|D,0\rangle,|D,1\rangle\}$ providing the strongest beneficial two-state sector among those shown.
{\bf b,} Fraction \(\eta_d=\lambda_{\max}(\Pi_d\mathcal G_R\Pi_d)/\lambda_{\max}(\mathcal G_R)\) of the unrestricted optimal weak-loss response captured by nested diagnostic subspaces of dimension \(d\), where \(\Pi_d\) projects onto the corresponding \(d\)-dimensional subspace. The minimal donor pair already captures \(82.5\%\) of the unrestricted optimum, increasing to \(95.7\%\) for three states and \(99.0\%\) for four states, and rapidly approaching unity as the subspace is enlarged.
{\bf c,} Physical realization of the resulting two-state admissible environment. Passive loss couples coherently to $\mathcal H_{\rm adm}=\mathrm{span}\{|D,0\rangle,|D,1\rangle\}$ through the normalized mode $|\varphi(q,\phi)\rangle=(|D,0\rangle+qe^{i\phi}|D,1\rangle)/\sqrt{1+q^2}$, with $K_\varphi=|\varphi\rangle\langle\varphi|$.
{\bf d,} Dimensionless weak-loss response $\zeta_\varphi= J\langle\varphi|G_{\rm vib}|\varphi\rangle/R_0^{\rm opt}$ within this admissible subspace. Although both local loss directions and every incoherent mixture are detrimental, their coherent superposition opens a positive-response region. The star marks the optimal mode $(q^*,\phi^*)=(0.347,0.626)$, where $\zeta_{\max}=0.140$.
{\bf e,} At fixed $\phi=\phi^*$, decomposition $\zeta_\varphi = \zeta_{\rm loc} + \zeta_{\rm coh}$. The local contribution is negative, whereas the coherent cross contribution is positive and exceeds the local loss penalty near $q=q^*$.
{\bf f,} Finite-loss consequence of the lossless-trajectory design. $R_{\rm traj}$ fixes the environmental mode at $(q^*,\phi^*)$ and reoptimizes only $(\nu,t)$, reaching an $8.73\%$ enhancement over the optimized lossless benchmark. Full finite-loss redesign gives $R_{\rm coh}^{\rm opt} /R_0^{\rm opt} =1.0952$, a $9.52\%$ enhancement, whereas numerical optimization finds that the incoherent family $R_{\rm inc}^{\rm opt}$ remains below the lossless benchmark throughout $0<\gamma/J\le4$. Panels a,b,d--f use $\Delta/J=1.65$ and $\kappa/J=1.20$.
}
\label{fig:Fig3}
\end{figure*}

\vspace{0.1cm}\noindent
{\bf Environmental coherence unlocks operational advantage}\\
The Bell protocol opens a beneficial channel by coherently coupling a detrimental bright direction to a neutral dark direction. More generally, both local loss directions may be detrimental. For two orthonormal admissible loss modes, the projected lossless-trajectory response takes the form \(G_2=\bigl(\begin{smallmatrix}-a&c\\c^*&-b\end{smallmatrix}\bigr)\), with \(a,b\geq0\).
Every incoherent mixture is then non-beneficial to first order, whereas a coherent superposition yields a positive response if and only if \(|c|^2>ab\). For \(a,b>0\), the cross response must therefore exceed a finite threshold to overcome both local loss penalties; the Bell protocol illustrates the special case with one neutral direction. This weak-loss criterion motivates the vibronic realization below, where finite-loss performance is tested separately.

We illustrate this principle with a minimal vibronic donor--acceptor model of vibration-assisted energy transfer~\cite{Chin2013,Irish2014,Dijkstra2015}, closely related to dynamics studied theoretically~\cite{Li2021,Li2022,Li2024} and implemented in trapped-ion quantum simulation~\cite{Gorman2018}. The model is governed by the Hamiltonian \(H_0 = \frac{\Delta}{2}\sigma_z + \frac{J}{2}\sigma_x + \nu a^\dagger a + \frac{\kappa}{2}\sigma_z(a+a^\dagger)\). 
Here \(\sigma_z= |A\rangle\langle A| -|D\rangle\langle D|\) and \(\sigma_x=|D\rangle\langle A|+|A\rangle\langle D|\) act on the donor--acceptor electronic subspace; \(\Delta\) and \(J\) are the donor--acceptor detuning and electronic coupling, respectively, while \(\nu\) and \(\kappa\) denote the vibrational frequency and electron--vibration coupling. The operator \(a\) annihilates a vibrational quantum, and \(|X,n\rangle =|X\rangle\otimes|n\rangle\) denotes a product state with electronic state \(X\in\{D,A\}\) and \(n=0,1,\ldots\) vibrational quanta. 
The system is initialized in \(|D,1\rangle\). The corresponding bare electronic splitting is \(\Omega_0 = \sqrt{\Delta^2+J^2}\).

The target operator is the projector onto the acceptor electronic manifold, \(A=P_A=\sum_n |A,n\rangle\langle A,n|\). Here, \(\rho(t)\) denotes the generally subnormalized density operator of the coupled electronic–vibrational system under passive loss, so that \(Y_A(t)=\operatorname{Tr}[P_A\rho(t)]\) gives the unconditional acceptor population at terminal time \(t\).
We therefore optimize the time-normalized target yield $R=Y_A(t)/t$ within the one-quantum vibronic-resonance operating manifold $\Theta_{\rm vib}$, corresponding to the branch near $\nu\simeq\Omega_0$ and its first target-transfer maximum. We define the independently optimized lossless benchmark \(R_{0}^{\rm opt} =\max_{(\nu,t)\in\Theta_{\rm vib}}R(\gamma=0;\nu,t)\), and denote its optimizer by \((\nu_0^*,t_0^*)\).

We first ask how much of the unrestricted dissipative opportunity can be captured by a physically minimal environment. Before imposing a particular admissible loss space, we define the unrestricted response operator \(\mathcal G_R = -(1/t_0^*)\int_0^{t_0^*} ds\,J_s\), evaluated along the optimized lossless vibronic trajectory at \((\nu_0^*,t_0^*)\). For any pair of vibronic states $|i\rangle$ and $|j\rangle$, let $\Pi_{ij}$ project onto their two-dimensional span and define the dimensionless pairwise response \(\zeta_{ij}^{(2)} =(J/R_0^{\rm opt})\lambda_{\max}(\Pi_{ij}\mathcal G_R\Pi_{ij})\). Positive $\zeta_{ij}^{(2)}$ therefore means that some coherent rank-one loss mode within that pair is beneficial to first order. 
Evaluating the lossless-trajectory response over the full vibronic Hilbert space reveals a strongly structured response: beneficial two-state sectors are concentrated among a small number of low-lying vibronic states, with the pair $\{|D,0\rangle,|D,1\rangle\}$ giving the largest positive pairwise response among the states shown in Fig.~\ref{fig:Fig3}a.

To quantify how rapidly the unrestricted opportunity is recovered in a low-dimensional sector, we project $\mathcal G_R$ onto a nested diagnostic sequence of low-lying vibronic subspaces, with projector $\Pi_d$ onto the corresponding $d$-dimensional subspace, and define the captured fraction \(\eta_d = \lambda_{\max}(\Pi_d\mathcal G_R\Pi_d)/\lambda_{\max}(\mathcal G_R)\). The two-state donor sector alone captures $82.5\%$ of the unrestricted maximum weak-loss response; the captured fraction rises to $95.7\%$ for three states, $99.0\%$ for four, and rapidly approaches unity thereafter (Fig.~\ref{fig:Fig3}b; Supplementary Note~IV).

This response analysis motivates the minimal admissible space $\mathcal H_{\rm adm} = \operatorname{span}\{|D,0\rangle,|D,1\rangle\}$, within which the environment couples to the normalized coherent loss mode \(|\varphi(q,\phi)\rangle = (|D,0\rangle+qe^{i\phi}|D,1\rangle)/\sqrt{1+q^2}\). 
Here ``coherent'' refers to the phase-coherent superposition of the two local loss couplings, not to population injection or recycling from the environment (Fig.~\ref{fig:Fig3}c).

At this optimum, for \(\Delta/J=1.65\) and \(\kappa/J=1.20\), projecting the unrestricted response onto $\mathcal H_{\rm adm}$ gives \(G_{\rm vib} = P_{\rm adm}\mathcal G_RP_{\rm adm}\), whose matrix elements in the ordered basis $\{|D,0\rangle,|D,1\rangle\}$ are \(G_{mn}=\langle D,m|G_{\rm vib}|D,n\rangle\), with \(G_{00}=-0.00402\), \(G_{11}=-0.20253\), and \(G_{01}=G_{10}^*=0.06345-0.04584i\). Both diagonal responses are negative, but the off-diagonal response satisfies \(|G_{01}|^2>G_{00}G_{11}\), ensuring a positive eigenvalue of \(G_{\rm vib}\).  
At weak loss, passive loss acting independently on either \(|D,0\rangle\) or \(|D,1\rangle\) is therefore detrimental, as is every normalized incoherent mixture \(K_{\rm inc}(p)=p|D,0\rangle\langle D,0|+(1-p)|D,1\rangle\langle D,1|\), with \(0\le p\le1\).

A coherent environmental coupling, however, accesses the off-diagonal response \(\langle\varphi|G_{\rm vib}|\varphi\rangle
= [G_{00}+q^2G_{11} + 2q\,{\rm Re}(e^{i\phi}G_{01})]/(1+q^2)\). Correspondingly, the dimensionless weak-loss response ($\zeta_\varphi= J\langle\varphi|G_{\rm vib}|\varphi\rangle/ R_{0}^{\rm opt}$) can be decomposed as $\zeta_\varphi =\zeta_{\rm loc}+\zeta_{\rm coh}$, with $\zeta_{\rm loc}=J(G_{00}+q^2G_{11})/ [(1+q^2)R_{0}^{\rm opt}]$ and $\zeta_{\rm coh}=2Jq\,{\rm Re}[e^{i\phi}G_{01}]/ [(1+q^2)R_{0}^{\rm opt}]$.
The local contribution $\zeta_{\rm loc}$ is strictly detrimental, whereas the coherent contribution $\zeta_{\rm coh}$ can overcome this cost (Fig.~\ref{fig:Fig3}d,e).
Diagonalization gives $\lambda_{\max}(G_{\rm vib})=0.02313$, with $(q^*,\phi^*)=(0.347,0.626)$ and maximum dimensionless weak-loss response $\zeta_{\max}=0.140$ (Fig.~\ref{fig:Fig3}d). 

We next test whether this weak-loss inverse design remains predictive at finite dissipation. At each loss strength, we fix the environmental mode \((q^*,\phi^*)=(0.347,0.626)\), identified entirely from the lossless trajectory, and reoptimize only the operating variables \((\nu,t)\) within \(\Theta_{\rm vib}\). This defines the trajectory-designed envelope \(R_{\rm traj}(\gamma)\), which reaches \(R_{\rm traj}/R_0^{\rm opt}=1.08734\) at \(\gamma/J=1.835\), an \(8.73\%\) increase (Fig.~\ref{fig:Fig3}f).
For comparison, at each fixed \(\gamma\) we additionally optimize the environmental mode \((q,\phi)\), defining the fully redesigned coherent envelope \(R_{\rm coh}^{\rm opt}(\gamma)\). Its maximum over \(\gamma\) reaches \(R_{\rm coh}^{\rm opt}/R_0^{\rm opt}=1.09525\) at \(\gamma/J=2.173\), corresponding to a \(9.52\%\) increase. Thus, in this realization, the mode inferred from linear response remains near-optimal well into the finite-loss regime. 
The finite advantage is likewise unavailable to the entire incoherent family. At each fixed \(\gamma\), independently optimizing the mixing weight \(p\) together with the same operating variables \((\nu,t)\) defines \(R_{\rm inc}^{\rm opt}(\gamma)\), which remains below \(R_0^{\rm opt}\) throughout \(0<\gamma/J\le4\) (Fig.~\ref{fig:Fig3}f).

The enhancement is unconditional: the environment supplies no population to the target and no survival conditioning is used. The phase-coherent passive-loss coupling reshapes interference among trajectory contributions, with the resulting target gain outweighing the direct penalty from population removal.

\begin{figure}[t]
\centering
 \includegraphics[width=1.\columnwidth]{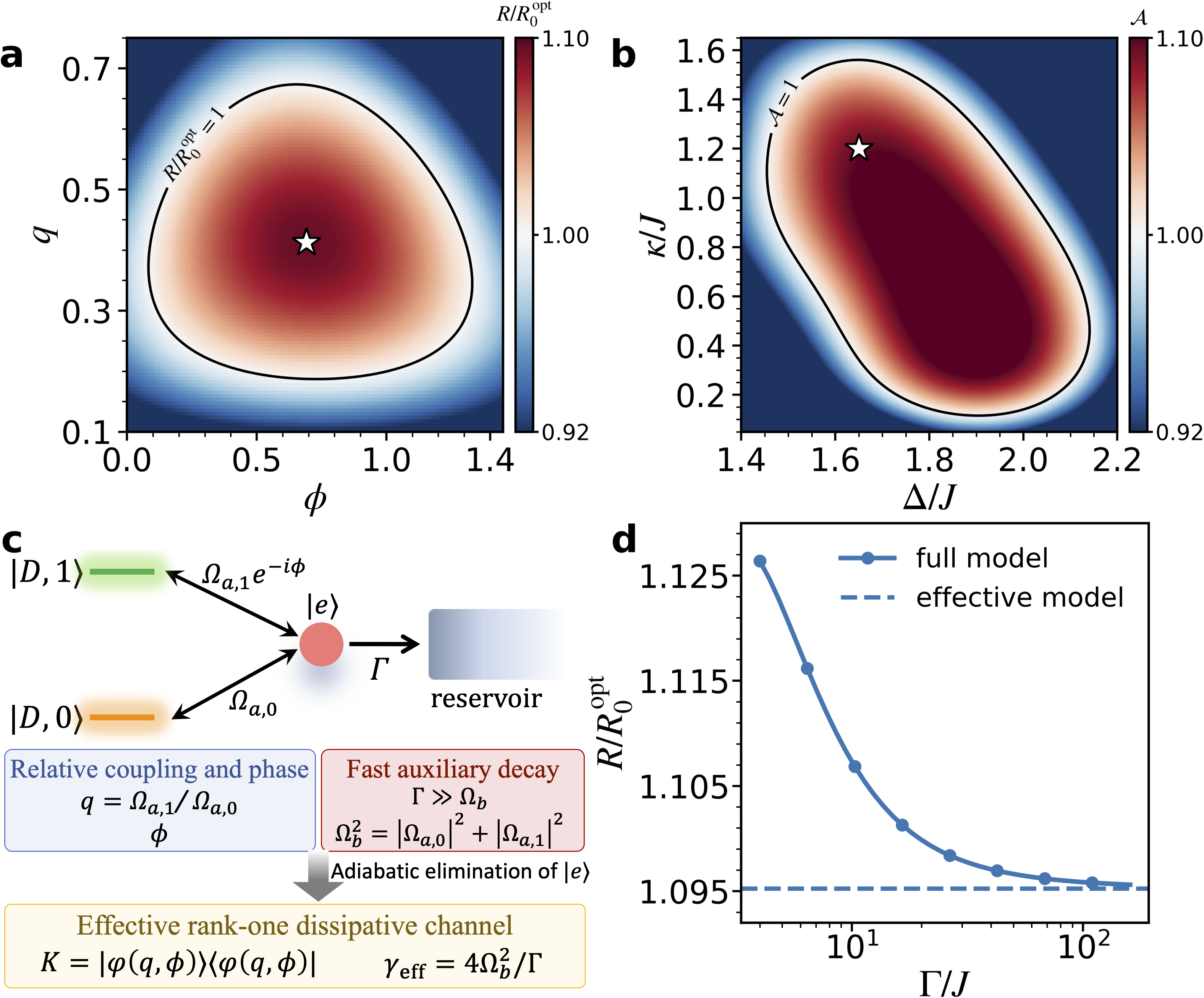} 
\caption{
{\bf Robustness and microscopic realization of beneficial passive dissipation.}
{\bf a,} Robustness to the environmental mode. The finite-loss operating point \((\gamma,\nu,t)\) is held fixed while the amplitude ratio \(q\) and relative phase \(\phi\) of the coherent environmental mode are varied. The black contour marks \(R/R_{0}^{\rm opt}=1\); the advantageous region inside it therefore occupies a finite neighbourhood rather than an isolated point. The star marks the finite-loss optimized mode \((q_{\rm opt},\phi_{\rm opt})=(0.413,0.692)\).
{\bf b,} Robustness to Hamiltonian parameters. The complete dissipative protocol is held fixed at its nominal optimum while the lossless reference is independently reoptimized within the corresponding vibronic operating manifold at each $(\Delta,\kappa)$. The black contour marks $\mathcal A=1$, where $\mathcal A=R^{\rm fixed}/R_0^{\rm opt}$ compares the target yield of this fixed dissipative protocol with the independently optimized lossless benchmark. The dissipative protocol remains advantageous over an extended neighbourhood of the nominal Hamiltonian. The star marks the nominal point $(\Delta/J,\kappa/J)=(1.65,1.20)$. 
{\bf c,} Microscopic realization of the coherent environmental mode. The donor-vibronic states $|D,0\rangle$ and $|D,1\rangle$ are coherently coupled to a rapidly decaying auxiliary state $|e\rangle$ with relative amplitude $q = \Omega_{a,1}/\Omega_{a,0}$ and phase $\phi$. On resonance and in the fast-decay regime $\Gamma\gg\Omega_b$, where $\Gamma$ is the auxiliary-state decay rate and $\Omega_b^2 = |\Omega_{a,0}|^2+|\Omega_{a,1}|^2$, adiabatic elimination generates the effective rank-one channel $K = |\varphi(q,\phi)\rangle \langle\varphi(q,\phi)|$ with $\gamma_{\rm eff} = 4\Omega_b^2/\Gamma$. 
{\bf d,} Full microscopic dynamics without adiabatic elimination. Keeping \(\gamma_{\rm eff}=2.173J\) fixed, with \((q,\phi,\nu,t)\) fixed at their finite-loss optimal values, the time-normalized target yield of the full system--auxiliary--reservoir model converges with increasing $\Gamma/J$ to the effective rank-one prediction $R/R_{0}^{\rm opt} =1.0952$. At $\Gamma/J=160$ the relative difference is approximately $3.2 \times 10^{-4}$.
}
\label{fig:Fig4}
\end{figure}

\vspace{0.1cm}\noindent
{\bf Robustness and microscopic realization}\\
For trajectory-based environmental design to be physically useful, the predicted advantage should not rely on exact parameter tuning and the coherent environmental mode should admit a microscopic realization.
Holding the finite-loss operating point fixed while varying the environmental mode gives a finite advantageous region around the optimized mode (Fig.~\ref{fig:Fig4}a). The advantage likewise persists over an extended neighbourhood of the nominal Hamiltonian when the complete dissipative protocol is held fixed at its nominal optimum, while the lossless reference is independently reoptimized at each perturbed Hamiltonian (Fig.~\ref{fig:Fig4}b). 

Finally, the coherent environmental mode can be generated from a conventional microscopic reservoir. On resonance, a rapidly decaying auxiliary state \(|e\rangle\), with decay rate \(\Gamma\), is coherently coupled to the two donor-vibronic states through \(H_{\rm aux} = |e\rangle \left(\Omega_{a,0}\langle D,0| + \Omega_{a,1}e^{-i\phi}\langle D,1| \right) +\mathrm{H.c.} \), where \(\Omega_{a,0}, \Omega_{a,1}>0\) denote the coupling magnitudes. 
Defining \(\Omega_b^2= | \Omega_{a,0}|^2 + |\Omega_{a,1}|^2\), adiabatic elimination in the fast-decay regime~\cite{Reiter2012} generates the designed rank-one operator \(K = |\varphi(q,\phi)\rangle \langle \varphi(q,\phi)|\), with \(q = \Omega_{a,1} / \Omega_{a,0}\) and \(\gamma_{\rm eff}=4\Omega_b^2/\Gamma\) on resonance (Fig.~\ref{fig:Fig4}c). The associated jump removes population from the retained vibronic manifold; the effective jump operator and derivation are given in Methods. 

We test this reduction directly by propagating the full system--auxiliary--reservoir dynamics while adjusting \(\Omega_b\) to keep \(\gamma_{\rm eff}=2.173J\) fixed. The target yield is calculated from the full density operator, without conditioning on survival or renormalizing the retained state. The microscopic value of \(R\) converges to the effective rank-one prediction as \(\Gamma\) increases (Fig.~\ref{fig:Fig4}d), with a relative difference of approximately \(3.2\times10^{-4}\) at \(\Gamma/J=160\).
Thus the enhancement persists in the full microscopic dynamics and is not an artefact of adiabatic elimination.

\vspace{0.1cm}\noindent
{\bf Discussion}\\
Sink-only removal exposes a distinction between classical and quantum dynamics: classical killing attenuates non-negative trajectory contributions, whereas quantum removal can reshape interference. Exploiting this difference requires a beneficial response that is physically accessible and accumulates along a fixed environmental mode. Beneficial passive loss is therefore a property of the target-dependent trajectory, rather than an instantaneous state.

Related effects include noise- and dephasing-assisted transport~\cite{Plenio2008,Mohseni2008,Rebentrost2009,ZerahHarush2018,Maier2019}, loss-induced transmission, revival and Zeno phenomena~\cite{Guo2009,Peng2014,Cerjan2016,Beder2024,Chen2021}, survival-conditioned non-Hermitian acceleration~\cite{Li2023,Zheng2025,Yu2026,Yuan2026}, and loss-controlled interference associated with non-Hermitian spectral structure~\cite{LiXue2026}. Purely dissipative enhancement of excitation transfer has also been reported, including release from transport-blocking dark subspaces~\cite{Caruso2009}. The present framework provides a lossless-trajectory criterion for beneficial sink-only removal that incorporates the unconditional survival cost, physical accessibility and trajectory accumulation, and identifies the optimal weak-loss mode directly from lossless dynamics.

In the vibronic realization, coherent environmental coupling combines two individually detrimental local loss directions into a beneficial mode, whereas every incoherent mixture remains detrimental to first order. This two-state sector captures most of the unrestricted weak-loss response, making the design accessible to a minimal reservoir. The mode selected from lossless dynamics remains near-optimal at finite loss, as verified by comparison with full finite-loss optimization. 

More broadly, the response operator \(G\) makes lossless dynamics a resource for environmental design~\cite{Koch2016,Gautier2025,Harrington2022}. For quantum tasks such as state preparation and excitation transfer, it provides a way to assess whether accessible passive loss should be suppressed or can instead be engineered to improve an
unconditional target.
By identifying the optimal weak-loss environmental mode, the criterion provides a principled starting point for finite-loss optimization. This establishes a predictive basis for reservoir engineering, linking the design of dissipation to the quantum dynamics that performs the intended task.


\vspace{0.5cm}\noindent
\noindent{\bf Methods}\label{methods}\\
\noindent{\bf Passive-loss response and inverse-design criterion}\\
We describe passive removal from the system of interest by
\begin{equation}
\dot{\rho}_{\gamma} = \mathcal L_0[\rho_{\gamma}]
-\frac{\gamma}{2}\{K,\rho_{\gamma}\},
\qquad K\geq0,
\label{eq:M_loss}
\end{equation}
where \(\mathcal L_0\) generates the lossless dynamics, \(\gamma\geq0\) sets the overall loss strength, and \(K\) specifies the environmental coupling. The anticommutator term describes irreversible population removal from the system of interest, with no recycling back into the retained system. 
For the Hamiltonian realizations considered below, \(\mathcal L_0[X]=-i[H_0,X]\), with \(\hbar=1\).

For a positive target operator \(A\), we use the unconditional target yield \(Y_A(\gamma,T) = \operatorname{Tr}[A\rho_\gamma(T)]\) and the time-normalized target yield \(R(\gamma,T) = Y_A(\gamma,T)/T\), as defined in the main text.
Linearization of Eq.~(\ref{eq:M_loss}) about the lossless dynamics gives the forward--backward response and the projected response \(G\) introduced in the main text; the full derivation is given in Supplementary Note~I.

We normalize admissible environmental couplings by \(K\geq0\), \(K=P_{\rm adm}KP_{\rm adm}\), and \(\operatorname{Tr}K=1\).
Although admissible \(K\) may have arbitrary rank, Hermiticity of \(G\) gives \(\max_K\operatorname{Tr}(KG)=\lambda_{\max}(G)\), attained by the rank-one projector \(K_{\rm opt}=|\varphi_{\rm opt}\rangle \langle\varphi_{\rm opt}|\), where \(|\varphi_{\rm opt}\rangle\) is a principal eigenvector of \(G\). 

For a fixed admissible loss channel \(K\), let \(\boldsymbol{\theta}\) denote the remaining control parameters, including the terminal time where applicable. 
At each loss strength \(\gamma\), we optimize over these parameters within the same admissible control family, defining \(R^{\rm opt}(\gamma;K) = \max_{\boldsymbol{\theta}} R(\gamma, \boldsymbol{\theta};K)\).
If the lossless optimum \(\boldsymbol{\theta}_0\) is a regular interior optimum, the envelope theorem gives \(\left.\frac{dR^{\rm opt}(\gamma;K)}{d\gamma} \right|_{0^+} = \operatorname{Tr} \!\left[K\,G(\boldsymbol{\theta}_0)\right]\). 
The independently optimized lossless trajectory therefore determines both whether beneficial passive dissipation is possible to first order and which admissible environmental mode maximizes the corresponding weak-loss response. Full derivations, including the forward--backward pathway representation and the relation between signed response and noncommutativity, are given in Supplementary Note~I.

\vspace{0.1cm}\noindent
{\bf Classical passive-loss monotonicity}\\
For a controlled classical Markov process, passive killing multiplies the non-negative terminal reward of each lossless trajectory by a survival factor in \([0,1]\). This gives the trajectory-wise ordering stated in the main text and, after taking the supremum over the common underlying control family and all admissible passive-loss profiles, the finite-strength bound \(R_{\rm cl}^{\rm opt}(\gamma)\leq R_{\rm cl}^{\rm opt}(0)\). The result assumes sink-like removal only, with no recycling, feedback, modification of the transition law of surviving trajectories, or conditioning on survival. The complete proof and scope are given in Supplementary Note~II.

\vspace{0.1cm}\noindent
{\bf Bell-state realization}\\
We consider the three-state subspace $\{|0\rangle,|10\rangle,|01\rangle\}$ with coherent Hamiltonian \(H_0=\Omega_1|10\rangle\langle0| + \Omega_2|01\rangle\langle0|+ \mathrm{H.c.}\) and Bell target $|T\rangle=(|10\rangle+|01\rangle)/\sqrt2$. 
Introducing \(|B\rangle = (\Omega_1|10\rangle + \Omega_2|01\rangle)/\Omega\) and \(|D\rangle = (\Omega_2|10\rangle - \Omega_1|01\rangle)/\Omega\)
where $\Omega=(\Omega_1^2+\Omega_2^2)^{1/2}$, gives $H_0=\Omega(|B\rangle\langle0|+|0\rangle\langle B|)$. The admissible loss space is $\mathcal H_{\rm adm}=\operatorname{span}\{|B\rangle,|D\rangle\}$.

For each fixed pair of coherent couplings \((\Omega_1,\Omega_2)\), the lossless time-normalized target yield \(R_0(T)=r^2\sin^2(\Omega T)/T\) is optimized only over the terminal time \(T\). The coupling asymmetry therefore specifies a constrained target--dynamics mismatch rather than an additional optimization variable. 
For the near-symmetry scan in Fig.~\ref{fig:Fig2}d, the total coupling \(\Omega=(\Omega_1^2+\Omega_2^2)^{1/2}\) is held fixed at unity while \(\epsilon\) is varied. 
The representative asymmetric protocol in Fig.~\ref{fig:Fig2}b,c uses \((\Omega_1,\Omega_2)=(1,0.4)\).
The analytical projected response, its trajectory integral, the cumulative response and the near-symmetry expansion are derived in Supplementary Note~III.

\vspace{0.1cm}\noindent
{\bf Vibronic model and optimization protocol}\\
The vibronic realization is described by 
\(H_0 = \frac{\Delta}{2}\sigma_z + \frac{J}{2}\sigma_x
+ \nu a^\dagger a + \frac{\kappa}{2}\sigma_z(a+a^\dagger)\), 
with initial state $|D,1\rangle$ and target operator projecting onto the acceptor electronic manifold. We use $J$ as the energy scale and define $\Omega_0 = \sqrt{\Delta^2+J^2}$. The nominal parameters are $\Delta/J=1.65$ and $\kappa/J=1.20$.

All comparisons are made within the same one-quantum vibronic-resonance operating manifold $\Theta_{\rm vib}$. This is the branch continuously connected to the resonance near $\nu\simeq\Omega_0$ together with the corresponding first target-transfer maximum in time. The lossless benchmark is defined independently as
\begin{equation}
R_{0}^{\rm opt} 
=
\max_{(\nu,t)\in\Theta_{\rm vib}}
R(\gamma=0;\nu,t).
\label{eq:M_vib_benchmark}
\end{equation}
At the nominal Hamiltonian parameters, we have \(\nu_0^*/\Omega_0 = 0.91485\), \(Jt_0^*=3.96420\), and \(R_{0}^{\rm opt}/J = 0.164761\). 
Every dissipative advantage reported here is therefore measured relative to an independently optimized lossless protocol within the same physical operating manifold.

For the minimal physical realization motivated by the unrestricted-response analysis, passive loss is restricted to $\mathcal H_{\rm adm} = \operatorname{span}\{|D,0\rangle,|D,1\rangle\}$. As shown in Fig.~\ref{fig:Fig3}a,b, this minimal donor sector captures $82.5\%$ of the unrestricted optimal weak-loss response; the complete pairwise and projected-subspace analysis is given in Supplementary Note~IV.
The coherent rank-one channel is \(K(q,\phi) = |\varphi(q,\phi)\rangle\langle\varphi(q,\phi)|\) where \(|\varphi(q,\phi)\rangle = (|D,0\rangle + qe^{i\phi}|D,1\rangle) / \sqrt{1+q^2}\),
and the effective vibronic state obeys
\begin{equation}
\dot\rho
=
-i[H_0,\rho]
-\frac{\gamma}{2}\{K(q,\phi),\rho\}.
\label{eq:M_vib_loss}
\end{equation}
The removed population is not recycled into the vibronic manifold.

Evaluating the response operator on the independently optimized lossless protocol identifies the principal environmental mode \((q^*,\phi^*)=(0.34688,0.62567)\), obtained entirely from the lossless trajectory. To test whether this weak-loss design remains predictive at finite loss, we fix this environmental mode and reoptimize only the operating variables within $\Theta_{\rm vib}$ at each fixed loss strength:
\begin{equation}
R_{\rm traj}(\gamma)
=
\max_{(\nu,t)\in\Theta_{\rm vib}}
R(\gamma,q^*,\phi^*;\nu,t).
\label{eq:M_Rtraj}
\end{equation}
For comparison, the fully optimized coherent envelope is
\begin{equation}
R_{\rm coh}^{\rm opt}(\gamma)
=
\max_{\substack{q,\phi\\
(\nu,t)\in\Theta_{\rm vib}}}
R(\gamma,q,\phi;\nu,t),
\label{eq:M_Rcoh}
\end{equation}
whereas the normalized incoherent channel \(K_{\rm inc}(p) = p|D,0\rangle\langle D,0| + (1-p)|D,1\rangle\langle D,1|\) with \(0\leq p\leq1\) defines
\begin{equation}
R_{\rm inc}^{\rm opt}(\gamma)
=
\max_{\substack{0\le p\le1\\
(\nu,t)\in\Theta_{\rm vib}}}
R(\gamma,K_{\rm inc}(p);\nu,t).
\label{eq:M_Rinc}
\end{equation}
Thus the three curves in Fig.~\ref{fig:Fig3}f differ only in the environmental degrees of freedom available for optimization, while the underlying operating manifold is identical.

The trajectory-designed curve reaches its maximum at $\gamma_{\rm traj}/J=1.83533$, with $\nu_{\rm traj}/\Omega_0=0.92070$ and $Jt_{\rm traj}=4.12981$. The fully optimized coherent protocol reaches its maximum at \(\gamma_{\rm opt}/J = 2.17304\), \((q_{\rm opt},\phi_{\rm opt})=(0.41251,0.69154)\), \(\nu_{\rm opt}/\Omega_0 = 0.92381\), and \(Jt_{\rm opt}=4.08578\). 
These parameters are used below in the robustness and microscopic realization tests. The finite-loss comparisons in Fig.~\ref{fig:Fig3}f are numerical results over the tested interval $0<\gamma/J\leq4$. Numerical implementation, branch tracking and convergence tests are given in Supplementary Note~IV.

\vspace{0.1cm}\noindent
{\bf Microscopic realization and numerical procedures}\\
The designed rank-one passive channel can be realized microscopically by coupling the admissible vibronic subspace to a rapidly decaying auxiliary level. Specifically, we introduce an auxiliary state $|e\rangle$, coupled to an external reservoir state $|r\rangle$, with
\( H_{\rm aux} = \Delta_e |e\rangle\langle e|
+ \left[\Omega_{a,0}|e\rangle\langle D,0|
+ \Omega_{a,1}e^{-i\phi}|e\rangle\langle D,1|
+\mathrm{H.c.} \right]\). 
The full system--auxiliary--reservoir dynamics is governed by 
\begin{equation}
\dot\rho_{\rm full}
=
-i[H_0+H_{\rm aux},\rho_{\rm full}]
+
\mathcal D[L_e]\rho_{\rm full} ,
\label{eq:M_full_aux}
\end{equation}
with irreversible decay \(L_e=\sqrt{\Gamma}\, |r\rangle\langle e|\), where \(H_0\) is the vibronic system Hamiltonian defined above and \(\mathcal D[L]\rho = L\rho L^\dagger -\frac12\{L^\dagger L,\rho\}\) denotes the Lindblad dissipator.

Defining \(\Omega_b^2 = |\Omega_{a,0}|^2 + |\Omega_{a,1}|^2\), the only vibronic superposition coupled to the auxiliary state is \(|B_{\rm aux}\rangle = (\Omega_{a,0}^*|D,0\rangle + \Omega_{a,1}^*e^{i\phi}|D,1\rangle)/\Omega_b\). For real positive coupling magnitudes, $q = \Omega_{a,1} / \Omega_{a,0}$ gives $|B_{\rm aux}\rangle=|\varphi(q,\phi)\rangle$.
The relative amplitude and phase of the microscopic couplings therefore directly implement the environmental mode selected by the trajectory response.

In the fast-decay regime, effective-operator elimination of the auxiliary level~\cite{Reiter2012} gives \(L_{\rm eff} = \sqrt{\Gamma}\,\Omega_b/(\Delta_e-i\Gamma/2) |r\rangle\langle B_{\rm aux}|\), with \(\gamma_{\rm eff} = \Gamma\Omega_b^2/[\Delta_e^2 + (\Gamma/2)^2]\) and \(\delta_{\rm eff} = -\Delta_e\Omega_b^2/[\Delta_e^2+(\Gamma/2)^2]\). 
Within the adiabatically eliminated description, the subnormalized effective vibronic state $\rho_{\rm eff}$ obeys
\begin{equation}
\dot\rho_{\rm eff} = -i[H_0+\delta_{\rm eff}K,\rho_{\rm eff}] 
- \frac{\gamma_{\rm eff}}{2}\{K,\rho_{\rm eff}\},
\label{eq:M_aux_reduced}
\end{equation}
where \(K=|B_{\rm aux}\rangle\langle B_{\rm aux}|\). 
On resonance $\Delta_e=0$, the coherent shift vanishes and \(\gamma_{\rm eff} = 4\Omega_b^2/\Gamma\) and \(K = |\varphi(q,\phi)\rangle\langle\varphi(q,\phi)|\), which reproduces the passive-loss model used in the main text.
At fixed resonant $\gamma_{\rm eff}$, \(\Omega_b/\Gamma = \frac12\sqrt{\gamma_{\rm eff}/\Gamma}\), so increasing $\Gamma$ systematically approaches the effective rank-one description. The complete elimination and its validity conditions are given in Supplementary Note~V.

In the full trace-preserving dynamics, population emitted through \(L_e\) is transferred to the external reservoir state. The target operator \(A=P_A= \sum_n|A,n\rangle \langle A,n|\), defined in the main text, acts on the retained vibronic manifold and is extended by zero on the auxiliary and reservoir states. Its expectation value therefore depends only on the retained block, \(Y_A^{\rm full}(t)= \operatorname{Tr}[A P_{\rm ret}\rho_{\rm full}(t)P_{\rm ret}]\), where \(P_{\rm ret}\) projects onto the retained vibronic manifold. The adiabatically eliminated state \(\rho_{\rm eff}(t)\) approximates
this block in the fast-decay regime, so \(Y_A^{\rm eff}(t)=\operatorname{Tr}[A \rho_{\rm eff}(t)]\) converges to \(Y_A^{\rm full}(t)\) in the controlled elimination limit.
Both yields are unconditional: neither involves normalization by the surviving population or postselection.

All vibronic calculations use $J$ as the energy scale and a vibrational cutoff $N_{\rm ph}=10$. At the nominal Hamiltonian, the one-quantum resonance branch is tracked numerically in the window \(0.80\leq\nu/\Omega_0\leq1.05\) and \(2.5\leq Jt\leq5.5\), which contains the continuously connected resonance and its first target-transfer maximum. These bounds are numerical search limits used to track the branch and do not define the physical operating manifold \(\Theta_{\rm vib}\). 
The lossless benchmark and dissipative protocols are optimized independently on this same branch.

For the environmental robustness map in Fig.~\ref{fig:Fig4}a, \((\gamma,\nu,t)\) are fixed at their finite-loss optimal values while \((q,\phi)\) are varied. For the Hamiltonian robustness map in Fig.~\ref{fig:Fig4}b, the complete dissipative protocol is held fixed at its nominal optimized values while \((\Delta,\kappa)\) are varied: \(\gamma/J\), \(q\), \(\phi\), \(Jt\), and the physical vibrational frequency \(\nu/J\) are all fixed. 
In particular, \(\nu/J\) is fixed rather than \(\nu/\Omega_0(\Delta)\). The lossless denominator is independently reoptimized within the corresponding \(\Theta_{\rm vib}(\Delta,\kappa)\) at every \((\Delta,\kappa)\), using the enlarged numerical tracking window \(0.78\le\nu/\Omega_0(\Delta)\le1.08\) and \(2.4\le Jt\le5.8\) to follow the same continuously connected one-quantum branch across the Hamiltonian scan. 
We denote the resulting target yield by \(R^{\rm fixed}(\Delta,\kappa)\); the robustness ratio is then \(\mathcal A(\Delta,\kappa) = R^{\rm fixed}(\Delta,\kappa)/R_{0}^{\rm opt}(\Delta,\kappa)\).

For the microscopic convergence test in Fig.~\ref{fig:Fig4}d, the full system--auxiliary--reservoir dynamics is propagated without adiabatic elimination. For each $\Gamma$, \(\Omega_b = \sqrt{\gamma_{\rm eff}\Gamma}/2\) is chosen so that $\gamma_{\rm eff}=2.173J$ remains fixed, while the environmental mode $(q,\phi)=(q_{\rm opt},\phi_{\rm opt})$ and operating point $(\nu,t)=(\nu_{\rm opt},t_{\rm opt})$ are held fixed at their finite-loss optimal values. 
The microscopic value of $R$ converges to the effective rank-one result with increasing $\Gamma/J$; at $\Gamma/J = 160$ their relative difference is approximately $3.2\times 10^{-4}$.

The lossless response is evaluated by direct time integration, and all finite-loss dynamics is propagated without renormalizing the retained state. Convergence with respect to the vibrational cutoff and response-integration grid, together with the complete robustness and microscopic-convergence data, is given in Supplementary Notes~IV--VI.
All target populations and time-normalized target yields reported here are unconditional quantities; no trajectory selection, survival renormalization or postselection is used.


\vspace{0.5cm}
\noindent\textbf{Author contributions}\\
Z.-Z.L. conceived the study, developed the theory, performed the numerical calculations and analyzed the results. Z.-Z.L. and X.X. interpreted and discussed the results and contributed to the development and writing of the manuscript.

\vspace{0.1cm}
\noindent
{\bf Funding}\\
This work was supported by the National Key Research and Development Program of China (Grant No.~2025YFE0217400) and the National Natural Science Foundation of China (Grant No.~12674619).

\vspace{0.1cm}
\noindent\textbf{Competing interests} \\
The authors declare no competing interests.\\

\vspace{0.1cm}
\noindent{\bf Data availability}\\
Source data are provided with this paper. Additional numerical data supporting the analyses are included in the Supplementary Information. Any further data supporting the findings of this study are available from the corresponding authors upon reasonable request.

\vspace{0.1cm}
\noindent{\bf Code availability}\\
The custom numerical code used to generate the results reported in this study is available from the corresponding author upon reasonable request.

\end{document}